\documentclass[draft,a4paper,12pt]{article} 
\usepackage{amsmath,amssymb,amsfonts, amsthm} 
\usepackage[mathscr]{eucal}
\usepackage{graphics} 
\usepackage{color, xcolor} 
\usepackage{comment}

\newtheorem{definition}{Definition}[section]
\newtheorem{assumption}{Assumption}[section]
\theoremstyle{definition}
\newtheorem{remark}{Remark}[section]

\newcommand{\envspace}{\vspace{2mm}}

\newcommand{\Real}{\mathcal{R}}

\newcommand{\PP}{\mathbb{P}} 
\newcommand{\QQ }{\mathbb{Q}}

\newcommand{\FF}{\mathcal{F}}

\newcommand{\EE}{\mathbb{E}}

\newcommand{\Xhat}{\widehat{X}}

\newcommand{\half}{\frac{1}{2}}

\newcommand{\Qc}{\mathcal{Q}} 
\newcommand{\ds}{\displaystyle}
\newcommand{\Xc}{\mathcal{X}} 
\newcommand{\Kc}{\mathcal{K}}
\newcommand{\Xh}{\widehat{X}} 
\newcommand{\Yc}{\mathcal{Y}}
\newcommand{\Yh}{\widehat{Y}} 
\newcommand{\Qh}{\widehat{\QQ}}
 
\newcommand{\Qr}{\QQ^R}

\newcommand{\sosd}{\succeq _2} 
\def\keywordname{{\bf Key words:}} 
\newcommand{\keywords}[1]{\par\addvspace\baselineskip\noindent\keywordname\enspace\ignorespaces #1}

 \title{Overview of utility-based valuation}

 \author{David German \\ \vspace{-2mm} {\small Claremont McKenna College} \\ \vspace{-2mm}{\small Department of Mathematics} \\ \vspace{-2mm}{\small 850 Columbia ave} \\ \vspace{-2mm}{\small Claremont, CA 91711, USA} \\ \vspace{-2mm}{\small phone: 1(909)607-7261} \\ \vspace{-2mm}{\small fax: 1(909)621-8419} \\{\small \texttt{dgerman@cmc.edu}}}

 \date{\today}

\begin{document}

\maketitle
\begin{abstract}
We review the utility-based valuation method for pricing derivative securities in incomplete markets. In particular, we review the practical approach to the utility-based pricing by the means of computing the first order expansion of marginal utility-based prices with respect to a “small” number of random endowments.

\end{abstract}

\section{Introduction}
\label{sec:intr}

The valuation of derivative securities by an economic agent represents a basic
problem of financial theory and practice. It is also one of the most studied
problems within various models.
In the framework of a complete financial model each contingent claim can be
replicated by a portfolio of traded securities. Therefore, it admits a
uniquely defined arbitrage-free price given as the initial wealth of such a
portfolio. While complete financial models have many computational advantages,
they are still only an idealistic representation (or approximation) of real
financial markets, as the exact replication of options is usually not
possible. Hence, the resulting arbitrage-free prices computed in these models
should be used in practice rather cautiously. Indeed, assume for a moment that
the illiquid contingent claims can suddenly be bought or sold at a price
$p^{trade}$ which only slightly differs from the price $p$ computed in a
complete financial model. The na\.ive interpretation of the price $p$ leads
the investor to take an infinite position in the contingent claims, which is,
clearly, nonsense from a practical point of view.

Due to
inability to replicate non-replicable derivative securities perfectly, the ownership of these
derivatives bears some risk. Therefore pricing in incomplete financial markets
becomes a non-trivial task. A classical approach in the economic theory is to
view the valuation of derivatives as a part of the problem of optimal
investment. Of course, in this case the resulting prices will depend not only
on the financial market of traded assets (as in arbitrage-free valuation
approach) but also on ``subjective'' characteristics of an economic agent such
as

\begin{itemize}
\item His or her risk preferences, which in the classical framework of
  Von Neumann-Morgenstern are specified by the reference probability
  measure $\PP$ and the utility function $U$ for consumption at
  maturity.
\item The current portfolio $(x, q)$ of the investor, where $x$ is the
  wealth invested into liquid securities, and $q=(q_i)_{1\le i\le m}$
  is the vector of his holdings in the non-traded contingent claims.
\item Investor's trading volume in the derivative securities.
\end{itemize}

The first item is required, since in our framework pricing is similar to
investment. The risk preferences specify the trade-off between risk and
return. Regarding the last two items note that in our utility-based valuation
framework the prices of non-replicable derivatives have non-linear dependence
on trading quantities, that is, the price of $q$ such securities each valued
at $p$ dollars is different from $qp$ dollars. In contrast to arbitrage-free
prices the utility based prices are not given by a single number. They
represent a function depending on the trading volume and the current position
of the investor.

In this paper we will give an overview of the utility-based valuation method and important properties of a risk-tolerance wealth process. Particular examples of computations of the prices in various incomplete market models can be found in the companion paper \cite{German:10b}.

\section{Model of financial market}
\label{sec:model-financ-mark}

Let us consider a model of a security market which consists of $d+1$ traded or
liquid assets: one zero coupon bond or a savings account with zero interest
rate, and $d$ stocks. We work in discounted terms, i.e. we suppose that the
price of the bond is constant, and denote by $S=(S^i)_{1\le i\le d}$ the price
process of the $d$ stocks. Stock prices are assumed to be semimartingales on a
filtered probability space $(\Omega, \FF, (\FF_t)_{0\le t<T}, \PP)$. Here $T$
is a finite time horizon, and $\FF=\FF_T$.

A (self-financing) portfolio is defined as a pair $(x,H)$, where the constant
$x$ represents the initial capital and $H=(H^i)_{1\leq i\leq d}$ is a
predictable $S$-integrable process, where $H_t^i$ specifies how many units of
asset $i$ are held in the portfolio at time $t$. The wealth process
$X=(X_t)_{0\leq t\leq T}$ of the portfolio evolves in time as the stochastic
integral of $H$ with respect to $S$:
\begin{equation}
  \label{eq:portfolio}
  X_t = x + \int_{0} ^{t} H_u dS_u,\quad 0\leq t\leq T.
\end{equation}
We denote by $\mathcal{X}(x)$ the family of wealth processes with non-negative
capital at any instant and with initial value equal to $x$:
\begin{equation}
  \label{eq:X(x)}
  \mathcal{X}(x) \triangleq \{X\geq 0: ~  X \mbox{ is defined by }
  (\ref{eq:portfolio})\}.
\end{equation}

A non-negative wealth process is said to be \emph{maximal} if its terminal
value cannot be dominated by that of any other non-negative wealth process
with the same initial value. In general, a wealth process $X$ is called
\emph{maximal} if it admits a representation of the form
\begin{displaymath}
  X =X'-X'', 
\end{displaymath}
where both $X'$ and $X''$ are non-negative maximal wealth processes. A wealth
process $X$ is said to be \emph{acceptable} if it admits a representation as
above, where both $X'$ and $X''$ are non-negative wealth processes and, in
addition, $X''$ is maximal. A paper \cite{DelbSch:97} by Delbaen and
Schachermayer contains many deep results on maximal and acceptable wealth
processes.

A probability measure $\mathbb{Q}\sim \mathbb{P}$ is called an \emph{\it
  equivalent local martingale measure} if any $X\in\mathcal{X}(1)$ is a local
martingale under $\mathbb{Q}$. The family of equivalent local martingale
measures is denoted by $\mathcal{Q}$. We assume throughout that
\begin{equation}
  \label{eq:NA}
  \mathcal{Q} \neq \emptyset.
\end{equation}
By the First Fundamental Theorem of Asset Pricing this condition is equivalent to
the absence of arbitrage opportunities in the model. A precise statement of
this important result is given in the seminal papers \cite{DelbSch:94} and
\cite{DelbSch:98} by Delbaen and Schachermayer. In particular, \eqref{eq:NA}
implies that a constant positive process is maximal.

In addition to the set of traded securities we consider a family of $m$
non-traded European contingent claims with payment functions $f=(f_i)_{1\le i
  \le m}$, which are $\FF$-measurable random variables, with maturity $T$. We
assume that this family is dominated by the terminal value of some
non-negative wealth process $X$, that is
\begin{equation*}
  \parallel f \parallel \triangleq \sqrt{\sum_{i=1}^m f_i^2} \le X_T,
\end{equation*}
which is also equivalent (see \cite{DelbSch:94}, Theorem 5.7) to the following
integrability condition
\begin{equation*}
  \sup_{\QQ\in\mathcal{Q}}\EE_{\QQ}[\parallel f \parallel ] < \infty.
\end{equation*}

We are interested in the case when it is not possible to replicate (in
some appropriate way) these securities, and therefore the following
assumption is required.

\begin{assumption}[\cite{KramSirb:06b}]
  \label{as:a1}
  For any $q\in \Real^m$ such that $q\ne 0$, the random variable $\langle
  q,f\rangle \triangleq\ds\sum_{i=1}^m q_if_i$ is not replicable.
\end{assumption}

This assumption is made only for simplicity of notation. It does not restrict
generality.

Now let us consider an investor or an economic agent. The agent's preferences
are specified by the utility function $U$ for consumption at maturity. The
utility function
\begin{displaymath}
  U:(0,\infty) \rightarrow (-\infty, \infty),
\end{displaymath}
is assumed to be strictly concave, strictly increasing and continuously
differentiable, and to satisfy the Inada conditions:
\begin{equation}
  \label{eq:inada} 
  U'(0)=\lim_{x\rightarrow 0}U'(x)=\infty, \quad U'(\infty)=\lim_{x\rightarrow
    \infty}U'(x)=0.  
\end{equation}
In addition to these standard conditions, following \cite{KramSch:99}, it is
assumed that the asymptotic elasticity of $U$ is strictly less than $1$, that
is
\begin{equation}
  \limsup_{x\rightarrow\infty}\frac{xU'(x)}{U(x)}<1. \label{eq:asympt_el}
\end{equation}

In order to satisfy conditions of the theorems in \cite{KramSirb:06b} the
following assumption is made.
\begin{assumption}[\cite{KramSirb:06b}]
  The utility function $U$ is two times continuously differentiable on
  $(0,\infty)$ and its relative risk-aversion coefficient
  \begin{equation}
    A(x)\triangleq - \frac{xU''(x)}{U'(x)}, \qquad x>0,
  \end{equation}
  is uniformly bounded away from zero and infinity, that is, there are
  constants $c_1>0$ and $c_2<\infty$ such that
  \begin{equation}
    c_1<A(x)<c_2, \qquad x>0.\label{eq:A_bound}
  \end{equation}\label{as:a2}
\end{assumption}

In fact, this assumption implies both the Inada conditions (\ref{eq:inada})
and the condition on asymptotic elasticity (\ref{eq:asympt_el}). For further
discussion regarding this assumption and two times differentiability of $U$ in
general, see \cite{KramSirb:06}.

Assume that the initial portfolio of the investor has the form $(x,q)$, where
$x$ is the liquid capital invested in the savings account and {\it liquid}
stocks, and the vector $q$ represents the quantities of the {\it illiquid}
contingent claims in the portfolio. The liquid part of the portfolio will be
changing over time, while the illiquid part is fixed until maturity.

The goal of the investor is to maximize the expected utility of terminal
wealth. Given the portfolio $(x,q)$, the quantity $u(x,q)$ that allows to
distinguish between different portfolio configurations and trading strategies
is called an indirect utility and is defined as
\begin{equation}
  \label{eq:exp_ut}
  u(x,q)=\sup_{X\in \Xc(x,q)} \EE[U(X_T+\langle q,f\rangle )], \quad (x,q) \in
  \mathcal{K}, 
\end{equation}
where $\Xc(x,q)$ is the set of acceptable processes with initial capital $x$
whose terminal values dominate $-\langle q,f\rangle$, that is
\begin{equation*}
  \Xc(x,q) \triangleq \{X \ : \ X \hbox{ is acceptable, } X_0=x \hbox{ and }
  X_T+\langle q,f\rangle\ge 0 \} 
\end{equation*}
and $\mathcal{K}$ is the interior of the cone of points $(x,q)$ such that the
set $\mathcal{X}(x,q)$ is not empty, that is
\begin{equation*}
  \Kc \triangleq \hbox{int} \{ (x,q)\in \Real^{m+1} \ : \ \Xc(x,q) \ne \emptyset \}.
\end{equation*}

The problem of optimal investment with random endowment \eqref{eq:exp_ut} has
been carefully studied by Hugonnier and Kramkov in \cite{HugonKram:04}.  It
was shown there that under the conditions of no-arbitrage \eqref{eq:NA} and
the asymptotic elasticity \eqref{eq:asympt_el} the upper bound in
\eqref{eq:exp_ut} is attained provided that $u(x,q) < \infty$.

\section{Marginal-utility based prices}
\label{sec:marg-util-based}

We are interested in the problem of evaluation of non-traded contingent claims
$f=(f_i)_{1\leq i\leq m}$. We need to attach some meaning to the price of a
security that is not traded. We begin with an intuitive explanation, and then
will make it precise.  Intuitively, a ``price'' is defined as a "threshold"
$p=(p_i)_{1\le i\le m}$ such that the economic agent is willing to buy the
$i$th contingent claim at a price less than $p_i$, sell it at a price grater
that $p_i$, and do nothing at $p_i$.

To make the above description precise we need to introduce an order relation
in the space of portfolio configurations involving random endowments. In other
words, given two arbitrary portfolios $(x_i,q_i), \ i=1,2$, the investor
should be able to say that $(x_1,q_1)$ is ``better'' than (``worse'' than,
``equal'' to) $(x_2,q_2)$.  The classical approach of Financial Economics is
to define the preferences of the investor with respect to the future random
payoffs in terms of their expected utilities, i.e. using (\ref{eq:exp_ut}). In
this case, the ``quality'' of a portfolio $(x,q)$ is expressed as the maximal
expected utility $u(x,q)$, which can be achieved by investing the liquid
amount $x$ in the financial market according to the optimal trading strategy.
This leads us to the following definition.

\begin{definition}[\cite{KramSirb:06b}]
  Let $(x,q)\in\Kc$ be the initial portfolio of the agent. A vector
  $p\in\Real^m$ is called a {\it marginal utility based price} (for the
  contingent claims $f$) at $(x,q)$ if
  \begin{equation*}
    u(x,q)\ge u(x',q')
  \end{equation*}
  for any $(x',q')\in \Kc$ such that
  \begin{equation*}
    x+\langle q, p \rangle = x'+\langle q', p \rangle.
  \end{equation*}
\end{definition}

The interpretation of this definition is that the agent's holdings $q$ in $f$
are optimal in the model where the contingent claims can be traded at time
zero at the marginal utility based price $p$. In other words, given the
portfolio $(x,q)$ the investor will not trade the options at the price
$p(x,q)$.

\section{Davis price}
\label{sec:davis-price}

When the initial portfolio of the investor does not contain contingent claims,
i.e. the portfolio consists of liquid wealth $x$ only, and $q=0$, the marginal
utility based price
\begin{displaymath}
  p(x)\triangleq p(x,0)
\end{displaymath}
can often be computed explicitly. Note that such a price $p(x)$ specifies the
direction of trade (but not the optimal trading volume!). This case was
extensively studied in the literature. The common references include the paper
\cite{Rubin:76} by Rubinstein (in economic literature) and the papers
\cite{Davis:97} by Davis and \cite{HugonKramSch:05} by Hugonnier, Kramkov and
Schachermayer (in mathematical finance literature). The latter paper contains
precise mathematical conditions for the price $p(x)$ to be defined uniquely
and to satisfy the key formula \eqref{eq:p(x)} below.

Let $u(x)$ be a short notation for the value function in the case without
random endowments, that is
\begin{equation}
  u(x) \triangleq u(x,0)=\sup_{X\in\Xc(x)} \EE[U(X_T)], \quad x>0. \label{eq:u_x}
\end{equation}
We assume that
\begin{equation}
  u(x)<\infty \hbox{ for some } x>0.\label{eq:fin_u}
\end{equation}

An important role in the future analysis will be played by the {\it marginal
  utility of the terminal wealth of the optimal investment strategy}, that is
by the random variable $U'(\Xh_T(x))$, where $\Xh(x)$ is the solution to
(\ref{eq:u_x}). Note that it is often easier to compute $U'(\Xh_T(x))$, rather
than the terminal wealth $\Xh_T(x)$ itself.

Let $V(y)$ be the Legendre transform (the conjugate function) of the
investor's utility function $U(x)$ defined as
\begin{equation}
  \label{eq:V(y)}
  V(y) \triangleq \sup_{x>0} \{ U(x)-xy \}, \quad y>0.
\end{equation}
Consider the following dual optimization problem:
\begin{equation}
  v(y) \triangleq \inf_{Y\in\Yc(y)} [V(Y_T)], \quad y>0, \label{eq:v_y}
\end{equation}
where $\Yc(y)$ is the family of non-negative supermartingales $Y$ such that
$Y_0=y$ and $XY$ is a supermartingale for all $X\in \Xc(1)$. Note that
$\Yc(1)$ contains the density processes of all $\QQ\in \Qc$.

We remind the reader that $v(y)$ is the dual function in the case that
the investor initially has only liquid wealth $x$ (as defined in
\eqref{eq:u_x}). If the lower bound in (\ref{eq:v_y}) is attained,
then the process $\widehat Y(y)$ attaining the infimum is called the
dual minimizer. Note that in general the process $\widehat Y(y)/y$ is
not the density process of some martingale measure $\mathbb{Q}\in
\mathcal{Q}$. It was shown in \cite{KramSch:99} that under the
conditions of no-arbitrage \eqref{eq:NA}, asymptotic elasticity
\eqref{eq:asympt_el} and boundedness of value function
\eqref{eq:fin_u} the dual minimizer exists for any $y>0$.

Assume that the Lagrange multiplier $y$ is dual to the initial wealth $x$ in
the sense that
\begin{equation*}
  y = u'(x) \quad (\text{or, equivalently, }  x=-v'(y))
\end{equation*}
and that
\begin{equation}
  \label{eq:1}
  \frac{\Yh_T(y)}{y}=\frac{d\Qh(y)}{d\PP}
\end{equation}
for some $\Qh(y) \in \mathcal{Q}$. In this case the marginal utility price
$p(x)$ is given by the following risk neutral evaluation formula
\begin{equation}
  \label{eq:p(x)}
  p(x)=\EE_{\Qh(y)}[f].
\end{equation}
It was shown in \cite{HugonKramSch:05} (under the additional condition that
$S$ is locally bounded) that the condition \eqref{eq:1} is necessary and
sufficient for the marginal utility based price $p(x)$ to be uniquely defined
for \emph{any bounded} contingent claim $f$.

\section{Sensitivity analysis and risk-tolerance \\ wealth processes}
\label{sec:sens-analys-util}

Assume now that the investor can trade contingent claims at the initial time
at a price $p^{trade}$.  A very important question from the practical point of
view is what quantity $q=q(p^{trade})$ the investor should trade at the price
$p^{trade}$. If the initial portfolio of the economic agent consists
exclusively of liquid wealth $x$, the optimal (static) position $q(p^{trade})$
in the illiquid contingent claims can be computed (at least intuitively) using
marginal utility based prices $p(x,q)$ from the following ``equilibrium''
condition:
\begin{equation}
  \label{eq:equil}
  p^{trade}=p(x-\langle p^{trade}, q(p^{trade}) \rangle, q(p^{trade})). 
\end{equation}
This equation has a natural economic interpretation:
\begin{enumerate}
\item In order to acquire $q$ stocks at the price $p$ the investor needs to
  spend the cash amount
  \begin{displaymath}
    \langle p, q\rangle \triangleq \sum_{i=1}^m p_i q_i.
  \end{displaymath}
\item The position $(x,q)$ is optimal given an opportunity to trade
  derivatives at $p$ if and only if $p = p(x,q)$.
\end{enumerate}

The practical use of \eqref{eq:equil} is rather limited. In the literature
there are almost no explicit computations of $p(x,q)$ with $q$ different from
zero. As an exception we refer to the papers \cite{MusZar1:04} by Musiela and
Zariphopoulou and \cite{Hend:02} by Henderson where some explicit computations
are done for the case of exponential utilities.

If it is not possible to compute the price $p(x,q)$ explicitly, one may try to
compute a linear approximation of the price. That is a linear expansion of the
first order for ``small'' values of $\Delta x$ and $q$.
\begin{equation}
  \label{eq:p(x,q)_lin_appr}
  p(x+\Delta x,q) = p(x)+p'(x)\Delta x +D(x)q+o(|\Delta x|+\parallel q \parallel).
\end{equation}
Here $p'(x)=(p_i'(x))_{1\le i \le m}$ is an $m$-dimensional vector and
$D_{ij}(x)= \frac{\partial p_i}{\partial q_j}(x,0),$ is an $m \times m$ matrix
with ${1\le i,j \le m}$.

The detailed analysis of the linear approximation \eqref{eq:p(x,q)_lin_appr}
is given in the paper \cite{KramSirb:06b} by Kramkov and S\^{i}rbu. They compute
the sensitivity parameters $p'(x)$ and $D(x)$ under very general (essentially
minimal) assumptions.  In addition to the natural {\it quantitative} problem
of the computation of the vector $p'(x)$ and the matrix $D(x)$ for any family
of contingent claims $f$, Kramkov and S\^{i}rbu also study the important questions
of the {\it qualitative nature} such as

\begin{enumerate}
\item Are the marginal utility based prices computed at $q=0$ locally
  independent of the initial capital, that is, does
  \begin{equation*}
    p'(x)=0
  \end{equation*}
  hold true?

  This is an important property since, for example, the price in Black and
  Scholes model does not depend on the initial wealth of an economic agent.

\item Does the sensitivity matrix $D(x)$ have full rank, that is, does
  \begin{equation*}
    D(x)q=0 \hbox{ if and only if } q=0, \quad q \in \Real^m,
  \end{equation*}
  hold true?

  To illustrate this property consider a model with only one contingent claim
  $f$ and $D(x)=0$. Then the linear expansion of first order does not show any
  dependence of price on quantity $q$. At least second order expansion is
  required to see the dependence.

\item Is the sensitivity matrix $D(x)$ symmetric?

  Consider two claims $f_1$ and $f_2$ with the corresponding Davis prices $p_1(x)$
  and $p_2(x)$ respectively. Suppose it is possible to trade {\it only} in the
  claim $f_1$ at the price $p_1^{trade}$, such that $p_1(x)>p_1^{trade}$. Then
  the agent would buy $f_1$. Now, suppose that it is possible to trade {\it
    only} in the claim $f_2$ at the price $p_2^{trade}$, such that
  $p_2(x)>p_2^{trade}$. Similarly, the agent would buy $f_2$. Assume now that
  it is possible to trade both securities {\it simultaneously}. In this case
  it is not necessarily true that the agent should buy $f_1$ and buy
  $f_2$. Here is an example. Suppose $f_1=c+f_2$ for some constant $c$. If at
  some moment $p_1^{trade}-p_2^{trade}<c$, the agent would buy $f_1$ and sell
  $f_2$.

  It is very desirable, therefore, to be able to work with different
  groups of financial markets independently, i.e., to decompose a
  multidimensional problem into a sequence of one-dimensional
  problems. In other words, it is important to be able to find a
  family of contingent claims $h=(h)_{1\le i\le m}$, spanning the same
  space as the contingent claims $f=(f)_{1\le i\le m}$ and such that,
  for the contingent claims $h$, a change in the traded price of $h_i$
  will only determine the agent to take a position in the $i$-the
  claim alone. If the sensitivity matrix for $h$ is diagonal, then it
  is possible to decompose a multidimensional problem into a sequence
  of one-dimensional problems. Therefore we can work with different
  groups of financial markets independently in the first order if and
  only if the matrix $D(x)$ is symmetric (and therefore can be
  diagonalized.)

\item Is the sensitivity matrix $D(x)$ negative semi-definite, that is, does
  \begin{equation*}
    \langle q, D(x)q \rangle \le 0, \quad q \in \Real^m,
  \end{equation*}
  hold true?

  In the case of one contingent claim, this property simply means that $q(x)$
  has to have the same sign as $p(x)-p^{trade}$, which is again related to the
  correct direction of trade.

\item Is the linear approximation stable? Is it true that for any $p^{trade}$
  the linear approximation
  \begin{equation*}
    p^{trade}\approx p(x)-\langle p^{trade}, q(x)\rangle p'(x)+D(x)q
  \end{equation*}
  of the ``equilibrium'' equation
  \begin{equation*}
    p^{trade}=p(x-\langle p^{trade}, q(x) \rangle, q(x))
  \end{equation*}
  has the ``correct'' solution? For example, in one-dimensional case will we
  have the property that the solution $q(x)$ of the approximation equation is
  positive if and only if $p^{trade}$ is greater than $p(x)$?
\end{enumerate}

It is very interesting that there is one single property of financial markets
that is responsible for the positive answer to all of the above questions.

\begin{definition}[\cite{KramSirb:06b}]\label{def:risk-tolerance}
  Let $x>0$ and denote by $\widehat{X}(x)$ the solution to \eqref{eq:u_x}.
  The process $R(x)$ is called the risk-tolerance wealth process if it is
  maximal and
  \begin{equation}
    R_T(x)=-\frac{U'(\widehat{X}_T(x))}{U''(\widehat{X}_T(x))}.\label{eq:R}
  \end{equation}
\end{definition}

In other words, $R(x)$ is the replication process for the random payoff
defined in the right hand side of (\ref{eq:R}). Since we are in the framework
of incomplete markets, it is either possible to replicate this random payoff,
or not. The key message of \cite{KramSirb:06b}, see Theorems 8 and 9, is that
the financial models where the risk-tolerance wealth process exist are exactly
the models with ``good'' qualitative properties of the first order expansion
\eqref{eq:p(x,q)_lin_appr} for an \emph{arbitrary} family of contingent claims
$f$.

\section{Properties of risk-tolerance wealth \\ processes}
\label{sec:prop-risk-toler}

Provided that the risk-tolerance wealth process exists, its initial value is
\begin{equation}
  \label{eq:R_0(x)}
  R_0(x)=-\frac{u'(x)}{u''(x)}.
\end{equation}
Note that the expression on the right-hand side is the risk-tolerance of the
economic agent at initial time. This quantity can be extracted in practice
from the current mean-variance preferences of the investor.

Regarding the evolution of the risk-tolerance wealth process over time we note
that by Theorem 4 in \cite{KramSirb:06b}
\begin{equation}
  \label{eq:2}
  \frac{R(x)}{R_0(x)}=\widehat{X}'(x) \triangleq \lim_{\Delta x \rightarrow 0}
  \frac{\widehat{X}(x+\Delta x) - \widehat{X}(x)}{\Delta x}.  
\end{equation}
The intuitive understanding of the above formula is the following. Our
economic agent has initial wealth $x$ and invests it according to optimal
investment strategy $\widehat{X}(x)$ (optimal in the sense that $\Xhat(x)$ is
the solution of (\ref{eq:u_x})). If this investor is given a small additional
cash amount $\Delta x$, then the investor will use it according to the
investment strategy $\Xh(x+\Delta x) - \Xh(x)$, which for small $\Delta x$ is
proportional to $R(x)$ by \eqref{eq:2}. That is, $R(x)$ is the answer to a
simple question: "If you have an extra dollar, how would you invest it?"  In
practice, every investor/bank/mutual fund should be able to answer this
question.

The following heuristic argument explains the formulas \eqref{eq:R_0(x)} and
\eqref{eq:2}. Assume that the investor receives a small additional cash amount
$\Delta x$. Denote by $\Phi$ the terminal wealth of the strategy used for the
investment per dollar of $\Delta x$. As the initial capital of the strategy is
1 we have that
\begin{equation}\label{eq:phi_cond}
  \EE_{\Qh(y)}[\Phi]=1,
\end{equation}
where $\Qh(y)$ is the martingale measure defined in \eqref{eq:1} and
$y=u'(x)$. We want to choose the asset $\Phi$ independently of $\Delta x$ so
that
\begin{equation}
  \label{eq:heur_R} 
  \EE[U(\Xh_T(x)+\Delta x \Phi)]=u(x+\Delta x) + o((\Delta
  x)^2).
\end{equation}
The Taylor expansion of the left hand side of (\ref{eq:heur_R}) gives
\begin{align*}
  &\EE[U(\Xh_T(x)+\Delta x \Phi)] \\
  &= \EE[U(\Xh_T(x))] + \EE[U'(\Xh_T(x))\Delta x \Phi] + \half \EE
  [U''(\Xh_T(x))(\Delta x)^2 \Phi^2] + o((\Delta x)^2) \\
  & = u(x) + \Delta x\EE\left[y\frac{d\Qh(y)}{d\PP}\Phi\right]\\
  & \qquad\quad\ +\half(\Delta
  x)^2\EE\left[\frac{U''(\Xh_T(x))}{U'(\Xh_T(x))}y\frac{d\Qh(y)}{d\PP}\Phi^2\right]
  + o((\Delta x)^2) \\
  &= u(x) + u'(x)\Delta x - \half (\Delta x)^2 u'(x)
  \EE_{\Qh(y)}\left[\frac{\Phi^2}{R_T(x)}\right] + o((\Delta x)^2).
\end{align*}
Hence, we need to select $\Phi$ so that
\begin{displaymath}
  \EE_{\Qh(y)}\left[\frac{\Phi^2}{R_T(x)}\right] \rightarrow \min. 
\end{displaymath}
A quick way to show that the lower bound above is attained at
\begin{equation}
  \label{eq:optim_Phi}
  \Phi  = \frac{R_T(x)}{R_0(x)}
\end{equation}
is by using the change of num\'{e}raire technique. We select the risk-tolerance
wealth process $R(x)$ as the new num\'{e}raire and denote by $\Qr(x)$ the
corresponding martingale measure:
\begin{equation}
  \label{eq:bbR(x)}
  \frac{d\Qr(x)}{d\widehat{\mathbb{Q}}(y)} = \frac{R_T(x)}{R_0(x)}. 
\end{equation}
We have
\begin{align}\label{eq:Phi}
  \EE_{\Qh(y)}\left[\frac{\Phi^2}{R_T(x)}\right] =
  \frac1{R_0(x)} \EE_{\Qr(x)}\left[\left(\frac{\Phi
        R_0(x)}{R_T(x)}\right)^2\right] \leq \frac1{R_0(x)}, 
\end{align}
where in the last step we used the Cauchy inequality. On the other hand,
\begin{equation}\label{eq:risk-tol}
  \EE_{\Qh(y)}\left[\frac{\left(\frac{R_T(x)}{R_0(x)}\right)^2}{R_T(x)}\right]=
  \frac1{R_0(x)}\EE_{\Qh(y)}\left[\frac{R_T(x)}{R_0(x)}\right] =\frac1{R_0(x)}.
\end{equation}
Therefore by comparing \eqref{eq:Phi} and \eqref{eq:risk-tol} we obtain
\begin{equation*}
  \EE_{\Qh(y)}\left[\frac{\Phi^2}{R_T(x)}\right]\le \EE_{\Qh(y)}\left[\frac{\left(\frac{R_T(x)}{R_0(x)}\right)^2}{R_T(x)}\right]
\end{equation*}
for any $\Phi$ satisfying \eqref{eq:phi_cond}.  This proves the
optimality of the choice \eqref{eq:optim_Phi} for $\Phi$ and, hence,
the formula \eqref{eq:2} for $R(x)$.

Once the optimality of \eqref{eq:optim_Phi} has been established we can
compare the second order terms in the quadratic expansions of the left and
right parts of \eqref{eq:heur_R}. Direct computations show that this leads to
\eqref{eq:R_0(x)}.

\section{Existence of risk-tolerance wealth process}
\label{sec:second-order-stoch}

The incomplete markets that we consider consist of two main ingredients: the
investor's utility function, and a set of traded securities on a filtered
probability space. Theorem 8 in \cite{KramSirb:06b} gives the equivalence
conditions for existence of $R(x)$ and the required qualitative properties of
the sensitivity matrix $D(x)$. It is also interesting to know whether it is
possible to relax the market model and still get the required properties of
the matrix $D(x)$. Theorems 6 and 7 in \cite{KramSirb:06b} give the answer to
this question.
	
Let us recall that a market model is called complete if it is arbitrage-free
and every bounded non-negative contingent claim is replicable. The market is
complete if and only if the family of equivalent probability measures contains
only one element $\QQ\sim\PP$. It can be shown that if the financial model is
complete, then the risk-tolerance wealth process is well-defined for any
utility function $U$ such that (\ref{eq:fin_u}) and Assumption \ref{as:a2}
hold true (in a complete market every claim is replicable).

If the model is arbitrage-free but is incomplete then the family of equivalent
martingale measures $\mathcal{Q}$ contains an infinite number of elements.

\begin{definition}
  Let $f$ and $g$ be non-negative random variables on $(\Omega, \mathcal{F},
  \mathbb{P})$. $f$ is said to \emph{second order stochastically dominate} $g$
  ( $f\sosd g$) if
  \begin{displaymath}
    \int_0^t \mathbb{P}(f\geq x)dx\geq \int _0^t
    \mathbb{P}(g \geq x)dx, \; t\geq 0.   
  \end{displaymath} 
\end{definition}
\begin{remark}
  It is well-known that $f\sosd g$ if and only if
  \begin{displaymath}
    \mathbb{E}[\phi(f)] \leq \mathbb{E}[\phi(g)]
  \end{displaymath}
  for any function $\phi = \phi(x)$ on $[0, \infty)$ that is convex,
  decreasing, and such that the expected values above are well-defined.
\end{remark}

\begin{definition}\label{def:universal_minimal}
  Let $\mathcal{Q}$ be the family of measures equivalent to $\PP$. The
  probability measure $\Qh\in\mathcal{Q}$ is called \emph{the universal
    minimal martingale measure} if its Radon-Nikodym derivative dominates the
  Radon-Nikodym derivatives of other elements of $\mathcal{Q}$ in the sense of
  the second order stochastic dominance, that is,
  \begin{equation}
    \label{eq:3}
    \frac{d\Qh}{d\PP} \sosd
    \frac{d\QQ}{d\PP} \quad \text{ for all } \mathbb{Q}\in \mathcal{Q}. 
  \end{equation}
\end{definition}

Theorem 6 in \cite{KramSirb:06b} asserts that if there exists the universal
minimal martingale measure $\Qh\in\mathcal{Q}$, then for {\it any} utility
function $U$ satisfying (\ref{eq:fin_u}) and Assumption \ref{as:a2} the
process $R(x)$ is well-defined (qualitative properties of $p(x)$ hold true)
and vice versa. This also explains the wording of the Definition
\ref{def:universal_minimal}, since $\Qh$ solves the dual problem of optimal
investment \eqref{eq:v_y} for any utility function $U$ and any initial wealth
$x$.

A complimentary result is stated in Theorem 7 in \cite{KramSirb:06b}. If an
arbitrary financial model is considered, then the {\it only} utility functions
allowing the existence of the risk-tolerance wealth process are power
utilities and exponential utilities.

\section{Computation of $D(x)$}
\label{sec:computation-dx}

Remember, that the described theory was developed in order to compute the first
order correction to the price $p(x)$ due to the presence of non-replicable
assets in the portfolio. In addition to establishing an equivalence between
the existence of the process $R(x)$ and the desired properties of the matrix
$D(x)$, Theorem 8 in \cite{KramSirb:06b} provides the necessary machinery
required to compute the first order correction to the price $p(x)$.

Hereafter we assume that the risk-tolerance wealth process $R(x)$ is
well-defined. We choose $\frac{R(x)}{R_0(x)}=\widehat{X}'(x)$ to be a new
num\'eraire. Let
\begin{displaymath}
  f^R=\frac{fR_0(x)}{R_T(x)}
\end{displaymath}
be the discounted payoffs of the contingent claims and for any wealth process
$X$
\begin{displaymath}
  X^R=\frac{XR_0(x)}{R(x)}.
\end{displaymath}
be its wealth expressed in terms of the num\'{e}raire $\frac{R(x)}{R_0(x)}$.

Let $\Qr(x)$ be the probability measure on $(\Omega, \FF)$ whose Radon-Nikodym
derivative under $\PP$ is given by
\begin{equation}
  \label{eq:8}
  \frac{d\Qr(x)}{d\PP}=\frac{R_T(x)\Yh_T(y)}{R_0(x)y}, 
\end{equation}
where $y=u'(x)$, $\Xh(x)$ is the solution to \eqref{eq:u_x} and
$\Yh(y)$ is the solution to \eqref{eq:v_y}.  Note that the process
$\frac{R_t(x)\Yh_t(y)}{R_0(x)y}$ is a positive uniformly integrable
martingale starting at 1, and therefore it defines a density process
for the probability measure $\Qr(x)$. See Theorem 2.2 in
\cite{KramSch:99} for details. Note that $X^R$ is a supermartingale
under this measure. Under this measure the price process of the
contingent claim $f^R$ expressed in the number of units $R/R_0$ is
\begin{equation*}
  P^R_t=\EE_{\Qr(x)}[f^R\mid \FF_t].
\end{equation*}

Consider now the Kunita-Watanabe orthogonal decomposition of the price process
$P^R$ under $\Qr(x)$
\begin{equation}
  P^R_t=M_t+N_t, \quad N_0=0.\label{eq:decomp}
\end{equation}
One can think of this decomposition in the following way. The process $M$ is
an $R(x)/R_0(x)$-discounted wealth process, which represents the hedging
process. The process $N$ is a martingale under $\Qr(x)$, which is orthogonal
to all $R(x)/R_0(x)$ -discounted wealth processes. Therefore $N$ is the ``risk
process'' -- the part of the price process $P^R$ that cannot be hedged due to
the incompleteness of the market. Now, Theorem 8 in \cite{KramSirb:06b} gives
the explicit form of the price correction:
\begin{equation}
  \label{eq:D(x)}
  D^{ij}(x)=\frac{u''(x)}{u'(x)}\EE_{\Qr(x)}[N^i_TN^j_T],\quad 1\le i,j\le m,
\end{equation}
where $N$ is defined in (\ref{eq:decomp}).

\section{Further assumptions}
\label{sec:summary-assumptions}

For completeness we have to mention that Theorem 8 of
\cite{KramSirb:06b} holds true under the following technical
assumptions.

Following \cite{KramSirb:06} we call a $d$-dimensional semimartingale $R$

\noindent {\it sigma-bounded} if there is a strictly positive predictable
(one-dimensional) process $h$ such that the stochastic integral $\int h dR$ is
well-defined and is locally bounded.

\begin{assumption}[\cite{KramSirb:06b}]
  The price process of the traded securities discounted by the solution
  $\widehat{X}(x)$ to (\ref{eq:u_x}), that is the $d+1$-dimensional
  semimartingale
  \begin{equation}
    S^{\widehat{X}(x)} \triangleq \left( \frac{1}{\widehat{X}(x)},
      \frac{S}{\widehat{X}(x)} \right), \label{eq:disc_S} 
  \end{equation}
  is sigma-bounded.\label{as:a3}
\end{assumption}

We refer to \cite{KramSirb:06}, Theorem 3 for sufficient conditions that
ensure the validity of this assumption. In particular, this assumption is
satisfied if $S$ is a continuous process, or if the original (incomplete)
model can be extended to a complete one by adding a finite number of
securities.

To facilitate the formulation of the assumptions on the random endowments $f$,
we change the num\'eraire from the bond to the normalized optimal wealth
process $\widehat{X}(x)/x$ and denote by
\begin{equation*}
  g_i(x)\triangleq x\frac{f_i}{\widehat{X}_T(x)}, \qquad 1\le i\le m,
\end{equation*}
the payoffs of the European options discounted by $\widehat{X}(x)$.

Let $\widetilde\QQ(x)$ be a probability measure equivalent to $\PP$ such that
\begin{equation}
  \label{eq:measQ}
  \frac{d\widetilde\QQ(x)}{d\PP}\triangleq\frac{\Xh(x)\Yh(x)}{xy}, \ y=u'(x),
\end{equation}
and let $\mathbf{H}^2_0(\widetilde\QQ(x))$ be the space of square
integrable martingales with initial value $0$ under the measure
$\widetilde\QQ(x)$ defined in \eqref{eq:measQ}. Denote
\begin{equation*}
  \mathcal{M}^2(x) \triangleq \left\{ M\in \mathbf{H}^2_0(\widetilde\QQ(x)) \ : \ M=\int
    HdS^{\Xh(x)} \right\}, 
\end{equation*}
where $S^{\Xh(x)}$ was defined in (\ref{eq:disc_S}).

\begin{assumption}[\cite{KramSirb:06b}]
  There is a constant $c>0$ and a process $M\in \mathcal{M}^2(x)$ such that
  \begin{equation*}
    \sum_{i=1}^m \mid g_i(x)\mid \, \le c+M_T.
  \end{equation*}
  \label{as:a4}
\end{assumption}

 
\section{Remark on implementation in practice}
\label{sec:remark-impl-pract}

Let us consider an investor who is already trading on the market and uses his
proprietary model to find the risk-neutral pricing probability measure $\Qh$
and as the result he can compute the price $p(x)$. However, the investor's
pricing model is linear in the size of trade. The advantage of the previously
described theory is that it makes it possible to compute the price correction
to the {\it already computed} linear price $p(x)$ without changing anything
and with a rather minimal effort. The ingredients that are required are:
\begin{itemize} 
\item The investor's risk-neutral probability measure $\Qh$, which is already
  implemented by the investor.
\item The investor's relative risk aversion coefficient $\alpha(x)\triangleq
  -x\frac{u''(x)}{u'(x)}$, which can be deduced from the mean-variance
  preferences.
\item $R(x)/R_0(x)$, the investor's decision how to spend an "extra dollar".
\end{itemize}
Once we have these three ingredients, we can compute the matrix $D(x)$ using
formula \eqref{eq:D(x)}. Examples of such computations can be found in \cite{German:10b}.

\bibliographystyle{alpha}
\bibliography{../BIB/finance}
\end{document}